\documentclass[]{emulateapj}
\usepackage{amsmath}
\usepackage{enumerate}
\usepackage{times}
\usepackage{graphics, subfigure}
\usepackage{epsfig}
\usepackage{multirow}
\usepackage{ulem}
\usepackage{pdfpages}

\def\mysimlt{\mathrel{\hbox{\rlap{\hbox{\lower4pt\hbox{$\sim$}}}\hbox{$<$}}}}
\def\mysimgt{\mathrel{\hbox{\rlap{\hbox{\lower4pt\hbox{$\sim$}}}\hbox{$>$}}}}
\newcommand{\Mcore}{M_{\textrm{core}}}
\newcommand{\der}{\delta_{\rho}}
\newcommand{\Mlog}{M_{\textrm{log}}}
\newcommand*{\mysim}{\mathord{\sim}}
\newcommand*{\mylesssim}{\mathord{\lesssim}}
\newcommand*{\mygtrsim}{\mathord{\gtrsim}}
\newcommand*{\myapprox}{\mathord{\approx}}
\begin{document}

\title{Failure of a neutrino-driven explosion after core-collapse may lead to a thermonuclear supernova}

\author{Doron Kushnir\altaffilmark{1,2} and Boaz Katz\altaffilmark{3}} \altaffiltext{1}{Institute for Advanced Study, Einstein Drive, Princeton, NJ 08540, USA} \altaffiltext{2}{Corresponding author, kushnir@ias.edu} \altaffiltext{3}{Department of Particle Physics and Astrophysics, Weizmann Institute of Science, Rehovot 76100, Israel}

\begin{abstract}
We demonstrate that $\mysim10\,\textrm{s}$ after the core-collapse of a massive star, a thermonuclear explosion of the outer shells is possible for some (tuned) initial density and composition profiles, assuming that the neutrinos failed to explode the star. The explosion may lead to a successful supernova, as first suggested by Burbidge et al. We perform a series of one-dimensional (1D) calculations of collapsing massive stars with simplified initial density profiles (similar to the results of stellar evolution calculations) and various compositions (not similar to 1D stellar evolution calculations). We assume that the neutrinos escaped with a negligible effect on the outer layers, which inevitably collapse. As the shells collapse, they compress and heat up adiabatically, enhancing the rate of thermonuclear burning. In some cases, where significant shells of mixed helium and oxygen are present with pre-collapsed burning times of $\mylesssim100\,\textrm{s}$ ($\myapprox10$ times the free-fall time), a thermonuclear detonation wave is ignited, which unbinds the outer layers of the star, leading to a supernova. The energy released is small, $\mylesssim10^{50}\,\textrm{erg}$, and negligible amounts of synthesized material (including $^{56}$Ni) are ejected, implying that these 1D simulations are unlikely to represent typical core-collapse supernovae. However, they do serve as a proof of concept that the core-collapse-induced thermonuclear explosions are possible, and more realistic two-dimensional and three-dimensional simulations are within current computational capabilities.
\end{abstract}



\section{Introduction}
\label{sec:Introduction}

There is a strong evidence that type II supernovae are explosions of massive stars, involving the gravitational collapse of the stars' iron cores \citep[][]{BBFH,87a,smartt2009} and the ejection of the outer layers. It is widely thought that the explosion is obtained due to the deposition in the envelope of a small fraction ($\mysim1\%$) of the gravitational energy ($\mysim10^{53}\,\textrm{erg}$) released in neutrinos from the core, leading to the $\mysim10^{51}\,\textrm{erg}$ observed kinetic energy of the ejected material \citep[see][for reviews]{Bethe90,Janka2012,Burrows2013}. One-dimensional (1D) simulations indicate that the neutrinos do not deposit sufficient energy in the envelope to produce the typical $\mysim10^{51}\,\textrm{erg}$ kinetic energy. While some two-dimensional (2D) studies indicate robust explosions \citep[][]{Bruenn2013,Bruenn2014,Nakamura2014,Suwa2014} and some indicate failures or weak explosions \citep[][]{Takiwaki2014,Dolence2015}, these studies are affected by the assumption of rotational symmetry and by an inverse turbulent energy cascade, which, unlike many physical systems, tends to amplify energy on large scales. Therefore, three-dimensional (3D) studies are necessary to satisfactorily demonstrate the neutrino mechanism, but so far 3D studies have resulted in either failures or weak explosions \citep[][]{Takiwaki2014,Lentz2015,Melson2015a,Melson2015b}.

\citet[][]{BBFH} suggested a different mechanism for the explosion during core-collapse that does not involve the emitted neutrinos. They suggested that increased burning rates due to the adiabatic heating of the outer shells as they collapse leads to a thermonuclear explosion \citep[see also][]{Hoyle60,Fowler64}. This has the advantage of naturally producing $\mysim10^{51}\,\textrm{erg}$ from the thermonuclear burning (with a gain of $\mysim \textrm{MeV}/b$, where $b$ stands for baryon) of $\mysim M_{\odot}$ of light elements. Alternatively, a fraction of $\textrm{MeV}/b$ naturally explains a velocity scale of supernovae of thousands of $\textrm{km}\,\textrm{s}^{-1}$ that is more robustly observed than the kinetic energy. While this mechanism can operate only if the neutrinos failed to eject the envelope, it would still be possible to see the neutrinos as observed in SN1987A \citep[][]{87a}. A few 1D studies suggested that this mechanism does not lead to an explosion because the detonation wave is ignited in a supersonic in-falling flow \citep[][]{Colgate66,WW82,Bodenheimer}. While these studies are discouraging, they only demonstrate that some specific initial stellar profiles do not lead to thermonuclear explosions, and they do not prove that thermonuclear explosions are impossible for all profiles. We find it striking that so little effort has been dedicated to studying this mechanism, given the relatively low computational requirements to examine it \citep[see also][for a brief historical account of how the thermonuclear mechanism was left behind]{Burrows1988,Janka2012}.

In this paper we revisit the collapse-induced thermonuclear supernovae mechanism. In Sections~\ref{sec:1D simulations} we perform a series of 1D calculations of collapsing massive stars with simplified initial density profiles and various compositions, assuming that the neutrinos had a negligible effect on the outer layers. We demonstrate that $\mysim10\,\textrm{s}$ after the core-collapse of a massive star, a successful thermonuclear explosion of the outer shells is possible for some (tuned) initial density and composition profiles that includes a significant layer of He--O mixture. In Section~\ref{sec:simple} we use simple analytic arguments to explain the qualitative features of the numerical calculations. A summary of the results and conclusions is given in Section~\ref{sec:discussion}.

\section{1D Simulations}
\label{sec:1D simulations}

In this section we perform a series of 1D calculations of collapsing massive stars with simplified initial density profiles and various compositions, assuming that the neutrinos had a negligible effect on the outer layers. The initial profiles are described in Section~\ref{sec:initial 1D} and our numerical tools are described in Section~\ref{sec:simulation}. In Section~\ref{sec:example} we demonstrate that $\mysim10\,\textrm{s}$ after the core-collapse of a massive star, a successful thermonuclear explosion of the outer shells is possible for some initial density and composition profiles that include a significant layer of He--O mixture. The ignition process in this simulation is analyzed in Section~\ref{sec:ignition}. In Section~\ref{sec:sensitivity} we examine the sensitivity of our results to the assumed initial profile.

\subsection{Initial profiles}
\label{sec:initial 1D}
The first step is to define the pre-collapse stellar profiles. These profiles cannot be inferred from observations and require the calculation of the final stages of stellar evolution, which are poorly understood \citep[see, e.g,][]{Smith2014} and are therefore uncertain. Nevertheless, there are several physical constraints that are likely to hold.
\begin{enumerate}[a.]
\item The star contains a degenerate iron core with a mass slightly smaller than the Chandrasekhar mass.
\item The initial profile is in a hydrostatic equilibrium.
\item The profile is stable with a constant or rising entropy (per unit mass) as a function of radius.
\item The local thermonuclear burning time, $t_{\textrm{b}}$,  at any radius $r$ in the profile is much longer than the free-fall time, $t_{\textrm{ff}}$, where
\begin{equation}
t_{\textrm{b}}=\varepsilon/\dot Q,
\end{equation}
\begin{equation}
t_{\textrm{ff}}=\frac{r^{3/2}}{\sqrt{2GM(r)}},
\end{equation}
$\varepsilon$ is the internal energy (per unit mass), $\dot Q$ is the thermonuclear energy production rate (per unit mass), and $M(r)$ is the enclosed mass.
\end{enumerate}
We note that the demand for stability may be relaxed if the growth time of perturbations is much longer than the dynamical time, but this is beyond the scope of this work.
Based on these constraints, we adopt the following simple parameterized profile.
\begin{enumerate}
\item A fixed mass of $1.2\,M_{\odot}$ within $r<2\cdot10^{8}\,\textrm{cm}$ is assumed to have already collapsed at $t=0$ and is not simulated.
\item The hydrogen envelope is ignored and the temperature is set to zero ($10^5\,\textrm{K}$ in practice) at the profile's fixed outer radius of $3\cdot 10^{10}\,\textrm{cm}$.
\item To allow the shape and amplitude of the density profile to be varied, the profile is composed of two regions with an adjustable transition radius $r_{\textrm{break}}$. The inner region $2\cdot10^{8}\,\textrm{cm}<r<r_{\textrm{break}}$ has a constant entropy (per unit mass) and the outer region $r_{\textrm{break}}<r<3\cdot 10^{10}\,\textrm{cm}$ has a density profile,
\begin{equation}
\rho=\frac{\Mlog}{4\pi r^3},
\end{equation}
(equal mass $\Mlog$ per logarithmic radius interval). The requirement of hydrostatic equilibrium implies that the density, pressure, and temperature profiles are set (up to minor adjustments due to the composition) by two free parameters that are chosen as the inner density, $\rho_{i}\equiv\rho(r=2\cdot 10^8\,\textrm{cm})$, and total mass, $\Mcore$. The transition radius, $r_{\textrm{break}}$, is adjusted accordingly.
\item The composition of the explosive shell is a mixture of helium and oxygen. This mixture is placed at the outer parts of the profile at radii $r>r_{\textrm{base}}$ where the temperatures are sufficiently low such that the ratio between the local burning time and the free-fall time, $t_{\textrm{b}}/t_{\textrm{ff}}$, is larger than a fixed threshold $t_{\textrm{b},0}/t_{\textrm{ff},0}$. The value of $r_{\textrm{base}}$ is chosen such that this ratio is exactly $t_{\textrm{b},0}/t_{\textrm{ff},0}$. At lower radii, pure oxygen (where $T<2\cdot 10^9\,\textrm{K}$) and silicon (where $T>2\cdot 10^9\,\textrm{K}$) are placed, which have negligible burning during the simulation.
\end{enumerate}

The above prescription has four free parameters.
\begin{enumerate}
\item $\rho_i$ - the density at $2\cdot 10^8\,\textrm{cm}$.
\item $\Mcore$ - the enclosed mass within $3\cdot 10^{10}\,\textrm{cm}$.
\item $r_{\textrm{O}/\textrm{He}}$ - the ratio of the oxygen and helium mass fractions in the explosive shell.
\item $t_{\textrm{b},0}/t_{\textrm{ff},0}$ - the ratio between the burning time and the free-fall time at the base of the explosive shell $r_{\textrm{base}}$.
\end{enumerate}
The additional parameters $r_{\textrm{break}}$ and $\Mlog$ that enter the profile description are set by the choice of $\rho_i$ and $\Mcore$.

We note that significant shells of mixed He--O are not currently expected in non-rotating stellar evolution models. Nevertheless, stellar evolution calculations of rotating massive stars generally predict the existence of a mixed He--O shell \citep[][]{Heger2000,Hirschi2004,Heger2005,Hirschi2005,Hirschi2007,Yusof2013}.\footnote{The composition profiles of Heger et al. can be found in http://2sn.org/stellarevolution/}

\subsection{Collapse Simulations}
\label{sec:simulation}

To simulate the collapse we use the 1D, Lagrangian version of the VULCAN code \citep[for details; see][]{Livne1993IMT}, which solves the equations of reactive hydrodynamics with a 13 isotope alpha-chain reaction network (similar to the 13 isotope network supplied with FLASH with slightly updated rates for specific reactions, especially fixing a typo for the reaction $^{28}\textrm{Si}(\alpha,\gamma)^{32}\textrm{S}$, which reduced the reaction rate by a factor $\myapprox4$.). We use a sufficient resolution (typically $\myapprox10\,\textrm{km}$ for the initial profile) such that all of our results are converged to better than $\mysim1\,\%$. We also use the 1D hydrodynamic FLASH4.0 code with thermonuclear burning \citep[Eulerian, adaptive mesh refinement;][]{Fryxell2000,dubey2009flash}, with the same reaction network as in VULCAN, in order to verify that our results do not depend on the numerical scheme. False numerical ignition may occur if the burning time in a cell becomes shorter than the sound crossing time \citep{Kushnir2013}. To avoid this, we modified both codes to include a burning limiter that forces the burning time in any cell to be longer than the cell's sound crossing time by suppressing all burning rates with a constant factor whenever $t_{\textrm{sound}}>f t_{\textrm{burn}}$  with $f=0.1$ \citep[see][for a detailed description]{Kushnir2013}. The numerical convergence established below implies that the limiter does not modify the resulting profiles.

We assume that neutrinos emitted during the collapse of the inner core do not lead to an explosion and escape with a negligible effect on the outer layers. We also neglect the gravitational mass loss from the neutrino emission \citep[which may lead to a very weak explosion with kinetic energy $\mysim10^{47}\,\textrm{erg}$ if the thermonuclear explosion fails as well;][]{Lovegrove2013,Piro2013}. The layers below $r=2\cdot10^{8}\,\textrm{cm}$ are assumed to have already collapsed, and the initial pressure within this radius is set to zero. The pressure at the simulation inner boundary, $r=10^{8}\,\textrm{cm}$, is held at zero throughout the simulation. The mass of material that (freely) flows through the boundary is added to the original collapsed mass of $1.2\,M_{\odot}$ and is taken into account in the gravitational field. The results are insensitive to the details of the collapse of the inner parts due to the supersonic flow near the boundary that does not allow information to propagate outward to the outer shells where thermonuclear burning takes place. To verify this, we experimented with other schemes for the collapse of the inner parts (e.g., the inner numerical node constrained to free-fall motion until crossing $r=10^{8}\,\textrm{cm}$), and found negligible effects on our results.

For most of the range of the possible values of the free parameters $\rho_i,\,M_{\textrm{core}},\,r_{\textrm{He}/\textrm{O}}$, and $t_{\textrm{b},0}/t_{\textrm{ff},0}$, the thermonuclear burning does not release sufficient energy to unbind the star. However, there is a range of profiles with reasonable parameters for which successful explosions occur. Before discussing the full set of simulations that were performed (Section~\ref{sec:sensitivity}), we describe in Sections~\ref{sec:example} and~\ref{sec:ignition} one successful explosion. The fact that some 1D profiles lead to successful explosions serves as a proof of concept for the possibility of collapse-induced thermonuclear supernovae.

\subsection{Example of a successful explosion}
\label{sec:example}

\begin{figure}
\includegraphics[width=0.5\textwidth]{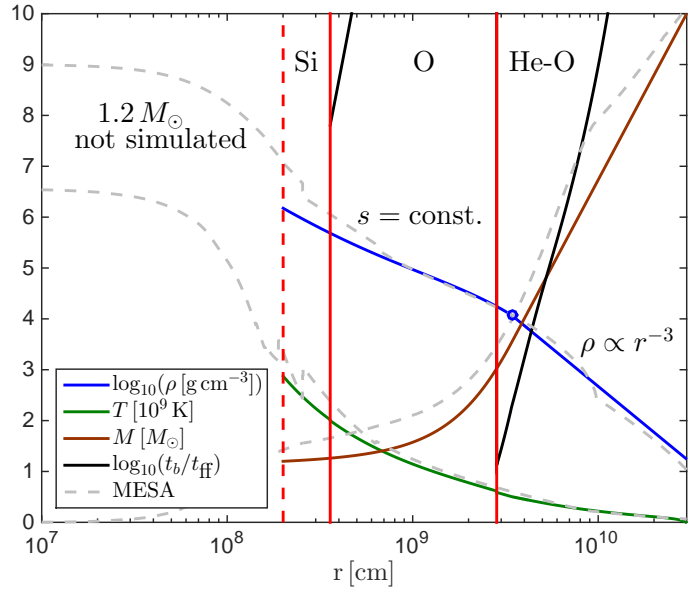}
\caption{Pre-collapse profile (density, temperature, enclosed mass, and burning to free-fall time ratio, $t_{\textrm{b}}/t_{\textrm{ff}}$) that leads to a successful explosion. The parameters for this profile are $\rho_i=1.5\cdot10^{6}\,\textrm{g}\,\textrm{cm}^{-3}$, $\Mcore=10\,M_{\odot}$ (leading to $r_{\textrm{break}}\approx3.42\cdot10^{9}\,\textrm{cm}$ and $\Mlog\approx3.0\,M_{\odot}$), $t_{\textrm{b},0}/t_{\textrm{ff},0}\approx13.3$, and $r_{\textrm{O}/\textrm{He}}=1$ (see Section~\ref{sec:initial 1D} for details). A fixed mass of $1.2\,M_{\odot}$ within $r<2\cdot10^{8}\,\textrm{cm}$ is assumed to have already collapsed at $t=0$ and is not simulated. The transition radius, $r_{\textrm{break}}\approx3.42\cdot10^{9}\,\textrm{cm}$, between constant entropy (per unit mass) and a density profile $\rho\propto r^{-3}$ is indicated with a blue circle. The base of the He--O mixture is at $r_{\textrm{base}}\approx2.84\cdot10^{9}\,\textrm{cm}$. At lower radii, pure oxygen (where $T<2\cdot 10^9\,\textrm{K}$) and silicon (where $T>2\cdot 10^9\,\textrm{K}$) are placed. For comparison, the pre-collapse profiles of a $30\,M_{\odot}$ star, calculated by Roni Waldman with the MESA stellar evolution code \citep[][]{MESA}, are shown (dashed gray).
\label{fig:InitialProfile}}
\end{figure}

A pre-collapse profile that leads to a successful explosion is shown in Figure~\ref{fig:InitialProfile}. The parameters for this profile are $\rho_i=1.5\cdot10^{6}\,\textrm{g}\,\textrm{cm}^{-3}$, $\Mcore=10\,M_{\odot}$ (leading to $r_{\textrm{break}}\approx3.42\cdot10^{9}\,\textrm{cm}$ and $\Mlog\approx3.0M_{\odot}$), $t_{\textrm{b},0}/t_{\textrm{ff},0}\approx13.3$ and a mixture of helium and oxygen with equal mass fractions $X_{\textrm{O}}=X_{\textrm{He}}=0.5$ ($r_{\textrm{O}/\textrm{He}}=1$). To achieve the required $t_{\textrm{b},0}/t_{\textrm{ff},0}$, the base of the He--O mixture is set to $r_{\textrm{base}}\approx2.84\cdot10^{9}\,\textrm{cm}$, with an enclosed mass of $m\approx3.03\,M_{\odot}$ (leading to $t_{\textrm{b},0}\approx71\,\textrm{s}$). The obtained density, temperature, and enclosed mass profiles are similar to the pre-collapse profiles of a $30\,M_{\odot}$ star, calculated by Roni Waldman with the MESA stellar evolution code \citep[][]{MESA}, which are shown for comparison. The main differences between the profiles are the existence and location of the He--O mixture.

\begin{figure}
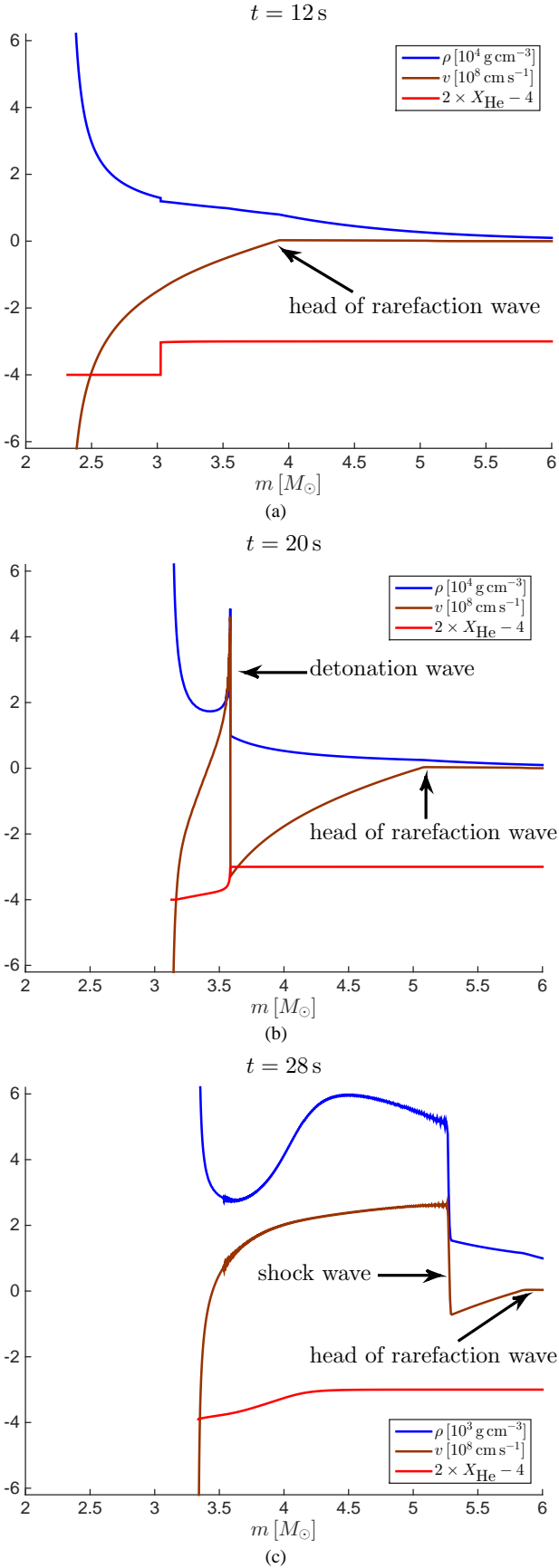

        \subfigure[]{
             \includegraphics[width=0.45\textwidth]{f2a.eps}
        }
        \subfigure[]{
             \includegraphics[width=0.45\textwidth]{f2b.eps}
        }
        \subfigure[]{
             \includegraphics[width=0.45\textwidth]{f2c.eps}
        }
      \caption{Dynamical evolution of the collapse, as calculated with VULCAN, for the initial conditions of Figure~\ref{fig:InitialProfile}. Each panel shows profiles (density, velocity, and $X_{\textrm{He}}$) for a snapshot of the simulation. Panel (a) $t=12\,\textrm{s}$; panel (b) $t=20\,\textrm{s}$; panel (c) $t=28\,\textrm{s}$. Note that the scale of the density is $10^{4}\,\textrm{g}\,\textrm{cm}^{-3}$ in panels (a) and (b), and is $10^{3}\,\textrm{g}\,\textrm{cm}^{-3}$ in panel (c).}
   \label{fig:Evolution}
\end{figure}

\begin{figure}
\includegraphics[width=0.5\textwidth]{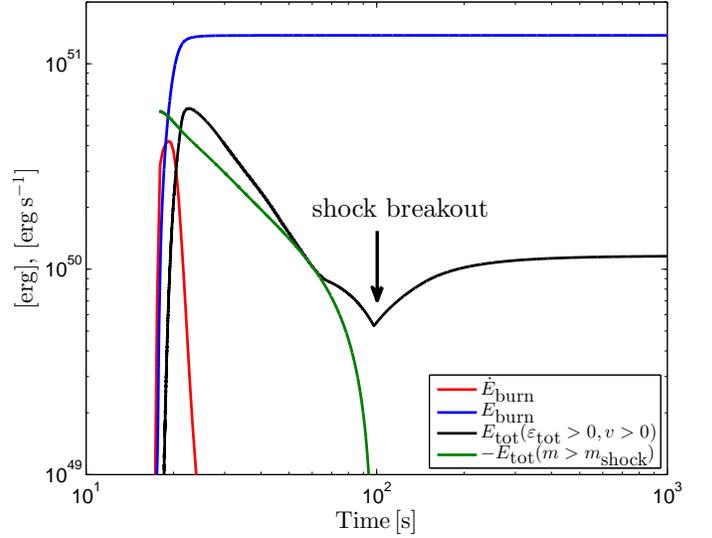}
\caption{Energy evolution during the collapse, as calculated with VULCAN, for the initial conditions of Figure~\ref{fig:InitialProfile}. The rate of thermonuclear energy production, $\dot{E}_{\textrm{burn}}$, is shown in red, and the accumulated thermonuclear energy produced, $E_{\textrm{burn}}$, is shown in blue. The total energy, including the gravitational, internal (not including potential thermonuclear), and kinetic energy of mass elements with positive velocity and positive total energy, $E_{\textrm{tot}}(\varepsilon_{\textrm{tot}}>0,\,v>0)$, is shown in black. The negative of the total energy of mass elements outward from the outgoing shock wave (or detonation wave), $-E_{\textrm{tot}}(m>m_{\textrm{shock}})$, is shown in green. Note that the last quantity is defined only after ignition, at $t\approx18\,\textrm{s}$.
\label{fig:Energetics}}
\end{figure}

The dynamical evolution of the collapse, as calculated with VULCAN, is shown in Figures~\ref{fig:Evolution} (snapshots from the simulation) and~\ref{fig:Energetics} (energy evolution) for the initial conditions of Figure~\ref{fig:InitialProfile}. A rarefaction wave propagates from the center of the star outward (evident as a velocity break appearing in panel (a) of Figure~\ref{fig:Evolution} at $m\approx3.95\,M_{\odot}$). Each element begins to fall inward as soon as the rarefaction wave reaches it. As it falls, each element is first slightly rarefied and then increasingly compressed. The velocity of the collapsing material increases and at some point the flow becomes supersonic. For example, $12\,\textrm{s}$ after the collapse the sonic point is located at $m\approx2.37\,M_{\odot}$. Sound waves cannot cross the sonic point outward, which is the cause for the low sensitivity to the exact inner boundary conditions, as explained above. As the base of the He--O shell is compressed and heated up adiabatically, the rate of thermonuclear burning is enhanced (which is the cause of the small density jump in panel (a) of Figure~\ref{fig:Evolution} at $m\approx3.03\,M_{\odot}$), and causes an ignition of a detonation wave at $t\approx18\,\textrm{s}$, as described in detail below. The ignition process takes place at a subsonic region (i.e., outward from the sonic point). An ignition of a detonation in a subsonic region occurred for all simulations in which a successful explosion was obtained.

The detonation wave propagates outward (panel (b) of Figure~\ref{fig:Evolution} at $m\approx3.6\,M_{\odot}$), producing thermonuclear energy at a rate of $\textrm{few}\times 10^{50}\,\textrm{erg}\,\textrm{s}^{-1}$ (Figure~\ref{fig:Energetics}). The pressure built from the accumulating thermonuclear energy manages to halt the inward collapse and cause an expansion that leads to an outward motion. Once the detonation wave reaches outer layers with densities $\rho\lesssim10^{4}\,\textrm{g}\,\textrm{cm}^{-3}$ it decays and transitions to a hydrodynamic shock that continues to propagate outwards (panel (c) of Figure~\ref{fig:Evolution} at $m\approx5.3\,M_{\odot}$). Note that the composition above the transition radius has a negligible effect on our results (and could be pure He, for example) as no further burning occurs. In this example, the shock reaches the stellar edge at $t\approx98\,\textrm{s}$ (Figure~\ref{fig:Energetics}), and the resulting ejecta has a mass of $\myapprox1.7M_{\odot}$ and a kinetic energy of $\myapprox10^{50}\,\textrm{erg}$. It is evident in Figure~\ref{fig:Energetics} that the potential energy of the burning shells and the mass external to them are of the same order as the released thermonuclear energy. The small kinetic energy of the ejecta is only a small fraction of the released thermonuclear energy of $10^{51}\,\textrm{erg}$. Furthermore, no post-collapse synthesized material is ejected. The properties of the ejecta may change if a hydrogen envelope is added.

\subsection{Ignition of a detonation}
\label{sec:ignition}

\begin{figure}
        \subfigure[]{
             \includegraphics[width=0.5\textwidth]{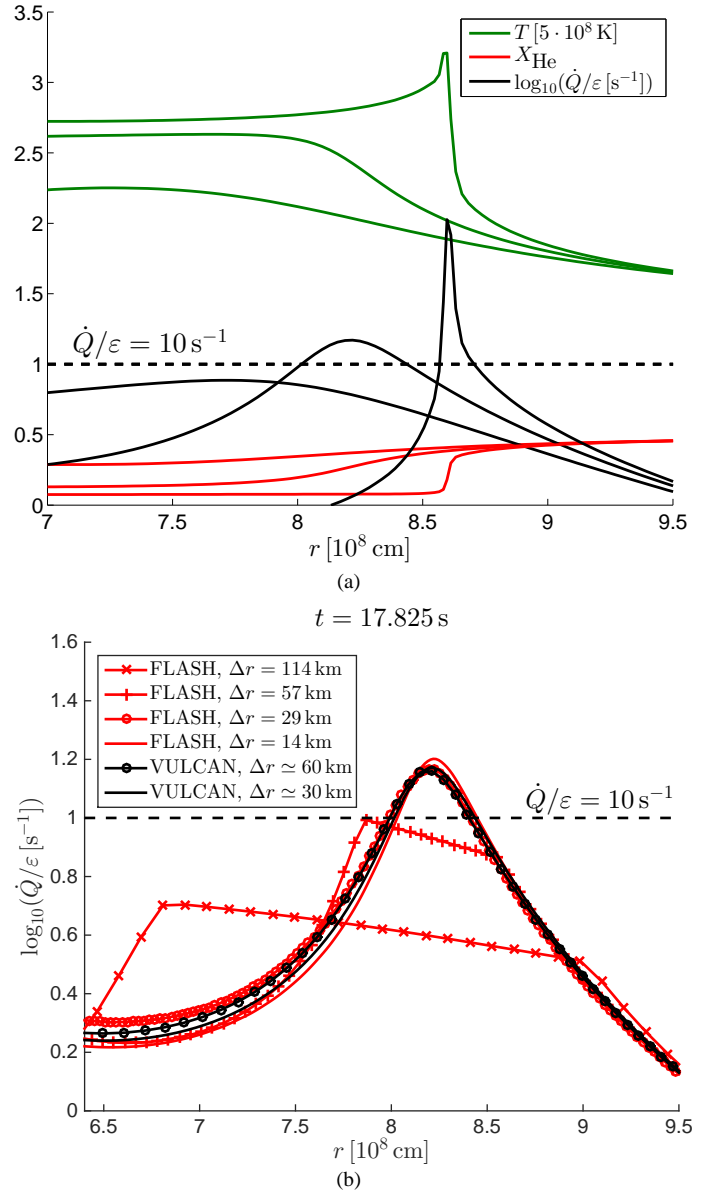}
        }
        \subfigure[]{
             \includegraphics[width=0.5\textwidth]{f4b.eps}
        }
      \caption{Ignition process for the initial conditions of Figure~\ref{fig:InitialProfile}. Panel (a): profiles (temperature, He mass fraction, and burning rate) from a VULCAN simulation at $t=17.7,\,17.825,\,17.95\,\textrm{s}$. The ignition of a detonation occurs when the burning rate becomes higher than the inverse sound crossing time $c_{\textrm{s}}/\Delta r\sim 10\,\textrm{s}^{-1}$ which is shown as a dashed black line, where $\Delta r\sim 5\cdot 10^{7}\,\textrm{cm}$ is the burning region size (as evident in the figure) and $c_{\textrm{s}}\approx 5\cdot10^{8}\,\textrm{cm}\,\textrm{s}^{-1}$ is the speed of sound (see the text for more details). The formation of the shock is clearly seen in $t=17.95\,\textrm{s}$. Panel (b): Snapshot of $\dot{Q}/\varepsilon$ at $t=17.825\,\textrm{s}$, in which the ignition criterion is met. VULCAN simulations (black) and FLASH simulations (red) are compared at different resolutions (for VULCAN, the actual resolution within the plotted region is given).
     }
   \label{fig:ignition}
\end{figure}

The ignition process in the example above is shown in Figure~\ref{fig:ignition}. For material near the base of the He--O shell, the collapse leads to a burning time that is comparable to the free-fall time at a radius of $\myapprox9.6\cdot10^{8}\,\textrm{cm}$, and He is efficiently consumed, leading to an ignition of a detonation. The condition for the formation of a detonation wave is that the thermal energy increases significantly (thereby increasing the burning rate) in a timescale shorter than the time it takes to hydrodynamically distribute the resulting excess pressure. The latter timescale is given by the sound crossing time $\Delta r/c_{\textrm{s}}$ of the burning region, where $c_{\textrm{s}}$ is the speed of sound and  $\Delta r\sim \dot{Q}/(d\dot{Q}/dr)$ is the length scale of the burning region. In case of a well defined burning wave, propagating with a phase velocity $v_{\varphi}$, and $\varepsilon\approx Q$, this condition reduces to the Zel'dovich criterion \citep[][]{zeldovich}, $v_{\varphi}>c_{\textrm{s}}$. The ignition condition is met at the time $t\approx18\,\textrm{s}$, shown in panel (a), where the scale of the burning region is $\Delta r\approx 5\cdot 10^{7}\,\textrm{cm}$, the typical speed of sound there is $\myapprox5\cdot10^{8}\,\textrm{cm}\,\textrm{s}^{-1}$, and the burning rate is $\dot{Q}/\varepsilon \mysimgt 10\, \textrm s^{-1}$. Note that at earlier times $\dot{Q}$ is significantly smaller, while $\Delta r$ is slightly larger, such that the ignition criterion is not met. Once the ignition criterion is met, significant thermonuclear energy is deposited locally, which increases the temperature and leads to a faster burning rate. This runaway process leads to the formation of a shock that is powered by the fast burning in its post shocked region, i.e., a detonation wave, as seen in $t=17.95\,\textrm{s}$. Because of the increased temperature and burning rate, the scale of the burning region decreases substantially, leading to the well known small length scale of thermonuclear detonation waves \citep[][]{Khokhlov89}. However, this small length scale is irrelevant to the ignition process \citep[contrary to what is commonly believed, e.g.,][]{Khokhlov89} which is determined at earlier times as explained here.

We are now in a position to estimate the numerical resolution required to resolve the ignition process. As seen in the snapshot at $t=17.825\,\textrm{s}$, in which the ignition criterion is met, the scale of the burning region is $\Delta r\approx 500\,\textrm{km}$, implying that a resolution of $\Delta r\sim 50\,\textrm{km}$ is sufficient to resolve the ignition process. Indeed, the VULCAN simulation is preformed with this resolution and $\dot{Q}/\varepsilon$ is converged to $\mysim 1\%$. A series of FLASH simulations with increasing resolutions is presented in panel (b) of Figure~\ref{fig:ignition}. As can be seen the (inverse) burning time, $\dot{Q}/\varepsilon$, is converged to a good approximation for resolutions $\Delta r\myapprox 10\,\textrm{km}$. This demonstrates that a modest resolution (that can be easily achieved in a full star simulation) is sufficient to resolve the ignition in this case. Note that at a slightly lower resolution ($\Delta r\approx 100\,\textrm{km}$) an ignition of a detonation is still obtained, although at a slightly different time and location. At much lower resolutions ($\Delta r\approx 500\,\textrm{km}$) an ignition of a detonation is not obtained.

\subsection{Successful 1D explosions require tuning}
\label{sec:sensitivity}

\begin{figure}
\includegraphics[width=0.5\textwidth]{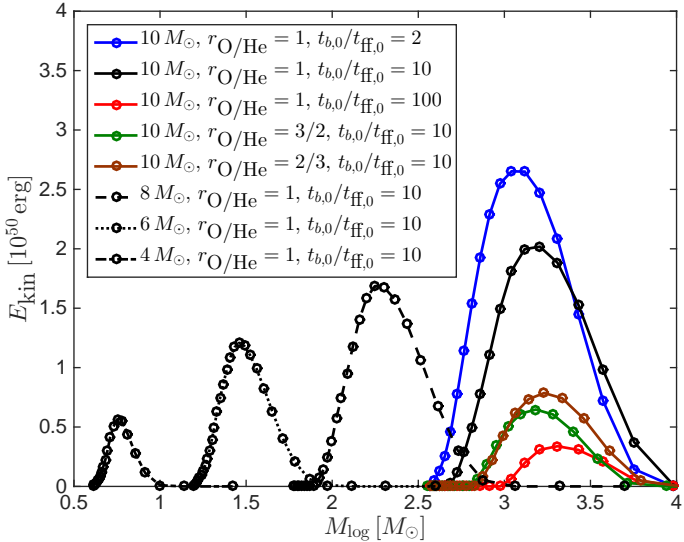}
\caption{Sensitivity of the asymptotic kinetic energy of the ejecta to the progenitor parameters (see Section~\ref{sec:initial 1D} for the description of the initial profiles and the free parameters). The dependence on the density normalization at the outer radii $r>r_{\textrm{break}}$, $\Mlog=dM(r)/d\ln(r)$, is shown on the $x$-axis, while the different curves probe the dependence on the other parameters, $\Mcore$, $r_{\textrm{O}/\textrm{He}}$, and $t_{\textrm{b},0}/t_{\textrm{ff},0}$ as indicated in the legend.
\label{fig:param}}
\end{figure}

In this section we examine the sensitivity of our results to the assumed initial profile. The asymptotic kinetic energy of the ejecta as a function of $\Mlog$ is shown in Figure~\ref{fig:param}. For $\Mcore=10$, $r_{\textrm{O}/\textrm{He}}=1$ and $t_{\textrm{b},0}/t_{\textrm{ff},0}=10$, asymptotic kinetic energy of $\mysim10^{50}\,\textrm{erg}$ is obtained for $2.9M_{\odot}\mysimlt \Mlog\mysimlt 3.55M_{\odot}$. For other values of $\Mlog$ the explosion fails (see below). We note that each simulation was checked for convergence. Increasing $t_{\textrm{b},0}/t_{\textrm{ff},0}$ to $100$ (i.e., increasing $r_{\textrm{base}}$) decreases significantly the asymptotic kinetic energy, and for higher values of $t_{\textrm{b},0}/t_{\textrm{ff},0}$ no explosions are obtained. Decreasing $t_{\textrm{b},0}/t_{\textrm{ff},0}$ to $2$ (much smaller values are not possible, since the shell burns before collapse) slightly increases the asymptotic kinetic energy. For initial compositions of $r_{\textrm{O}/\textrm{He}}=3/2$ and $r_{\textrm{O}/\textrm{He}}=2/3$, asymptotic kinetic energies that were lower by a factor of $\myapprox3$ were obtained, and for $r_{\textrm{O}/\textrm{He}}=7/3$ and $r_{\textrm{O}/\textrm{He}}=3/7$ the explosion fails (not shown in the figure). We also show the results for $\Mcore=4,6,8\,M_{\odot}$, in which smaller asymptotic kinetic energies are obtained for smaller $\Mcore$. In all cases, successful explosions requires burning times of $\mylesssim100\,\textrm{s}$. A list of the simulations in which the asymptotic kinetic energy of the ejecta is larger then $5\cdot10^{49}\,\textrm{erg}$ is in given in Table~\ref{tbl:list}.

In all cases in which the explosion fails a similar behavior is obtained. While the detonation wave is ignited and produces thermonuclear energy, similarly to the case studied in Section~\ref{sec:example}, the released thermonuclear energy is smaller than the potential energy of the burning shells and the mass external to them.

We also experimented with other compositions for the explosive shell (including protons, He, C, O, and heavier elements), and explosions were obtained only for He--O mixtures (with roughly equal mass fractions, as demonstrated above). This result can be understood as follows \citep[see also][]{Hoyle60}. The explosive shell cannot be composed only from protons, since the burning time of the mixture should be shorter than the dynamical time of the star and the p+p reaction is too slow (as it involves a weak reaction). In fact, the explosive shell cannot include a significant fraction of protons, despite the possibility of fast proton capture on heavier nuclei. The reason is that after a few proton captures the resulting nucleus will be too proton-rich for another capture, and the process must include a beta decay, which is too slow. For example, consider the reactions $^{12}$C$(p,\gamma)^{13}$N$(p,\gamma)^{14}$O. Since $^{15}$F is unstable against proton decay, further proton captures may proceed only after a beta decay of $^{14}$O with a half life of $\myapprox70\,\textrm{s}$, which is too long. One can imagine beginning with a roughly equal number of protons and heavier nuclei, but such a mixture is not energetic enough (see below). Another requirement is that the released thermonuclear energy is high enough to overcome the binding energy of the star, which requires a $\textrm{MeV}/b$ yield (see Section~\ref{sec:simple}). Mixtures that do not contain significant fractions of He cannot fulfill this requirement, as the energy per baryon decreases for heavier nuclei. These mixtures also typically ignite at high temperatures, when the material is already too deep in the potential well of the star, and the thermonuclear energy has a higher binding energy to overcome. We did not obtain explosions for pure helium shells, since the triple alpha reaction weakly depends on temperature for the relevant temperature range and the $\rho^2$ dependence of the reaction is not steep enough to allow ignition after a small amount of infall. These considerations leave only He--C and He--O as viable mixtures. It is hard to determine in advance which mixture is better, but our detailed simulations show that explosions can be obtained only for He--O mixtures. The reason is probably that for He--O mixtures the steep increase of the reaction rate with temperature happens at a slightly lower temperature than for He--C mixtures.

In summary, the required profiles are tuned and require the presence of a mixture of He and O with burning times of $\mylesssim100\,\textrm{s}$ ($\myapprox10$ times the free-fall time) prior to collapse, which is not currently expected in stellar evolution models. While the uncertainties involved with the pre-collapse final evolution stages of the star do not allow us to determine whether the initial conditions that lead to explosions in 1D are possible, they are probably unlikely.

\section{Approximate analytic treatment}
\label{sec:simple}

The numerical experiments described above imply that an explosion is possible, but only for a narrow range of initial profiles. In this section we attempt to provide an approximate analytic explanation for these results.

In successful explosions, a detonation wave is formed in a collapsing shell that propagates faster than the infall speed and manages to propagate out. As the wave traverses the progenitor, thermonuclear energy of order $\textrm{MeV}/b$ is released and (mostly) accumulated. At some point the wave reaches radii where the density is too low to support it and the thermonuclear burning is halted. In successful explosions, this energy is greater than the potential energy of the traversed shells and the mass external to them, which is of the order of $10^{51}\,\textrm{erg}$.

It is implied that there are two basic requirements for a successful thermonuclear explosion in a collapsing star,
\begin{enumerate}
\item an ignition of a detonation needs to occur at a sufficiently large radius so that the detonation wave propagates faster than the in-falling material; and
\item the detonation wave should traverse a significant amount of mass ($\mygtrsim M_{\odot}$) before it fades out in order to allow $\mysim 10^{51}\,\textrm{erg}$ to be released.
\end{enumerate}

The hydrodynamical collapse is analyzed in section \ref{sec:Collapse}. This allows the derivation of approximate conditions for the formation of an outgoing detonation wave. In addition, it is shown that at any given time, the amount of in-falling mass that is compressed to a density significantly higher than its initial (Lagrangian) density is very small $\ll \rm M_{\odot}$. This implies that in successful explosions, most of the contributing thermonuclear burning occurs in regions that have not suffered significant collapse. The approximate conditions for the thermonuclear burning of a significant amount of mass can therefore be found by analyzing the structure of the initial profile, assuming that a detonation wave traverses it. This is done in section \ref{sec:Explosion}.

\subsection{Collapse}\label{sec:Collapse}
\begin{figure}
\subfigure[]{
             \includegraphics[width=0.5\textwidth]{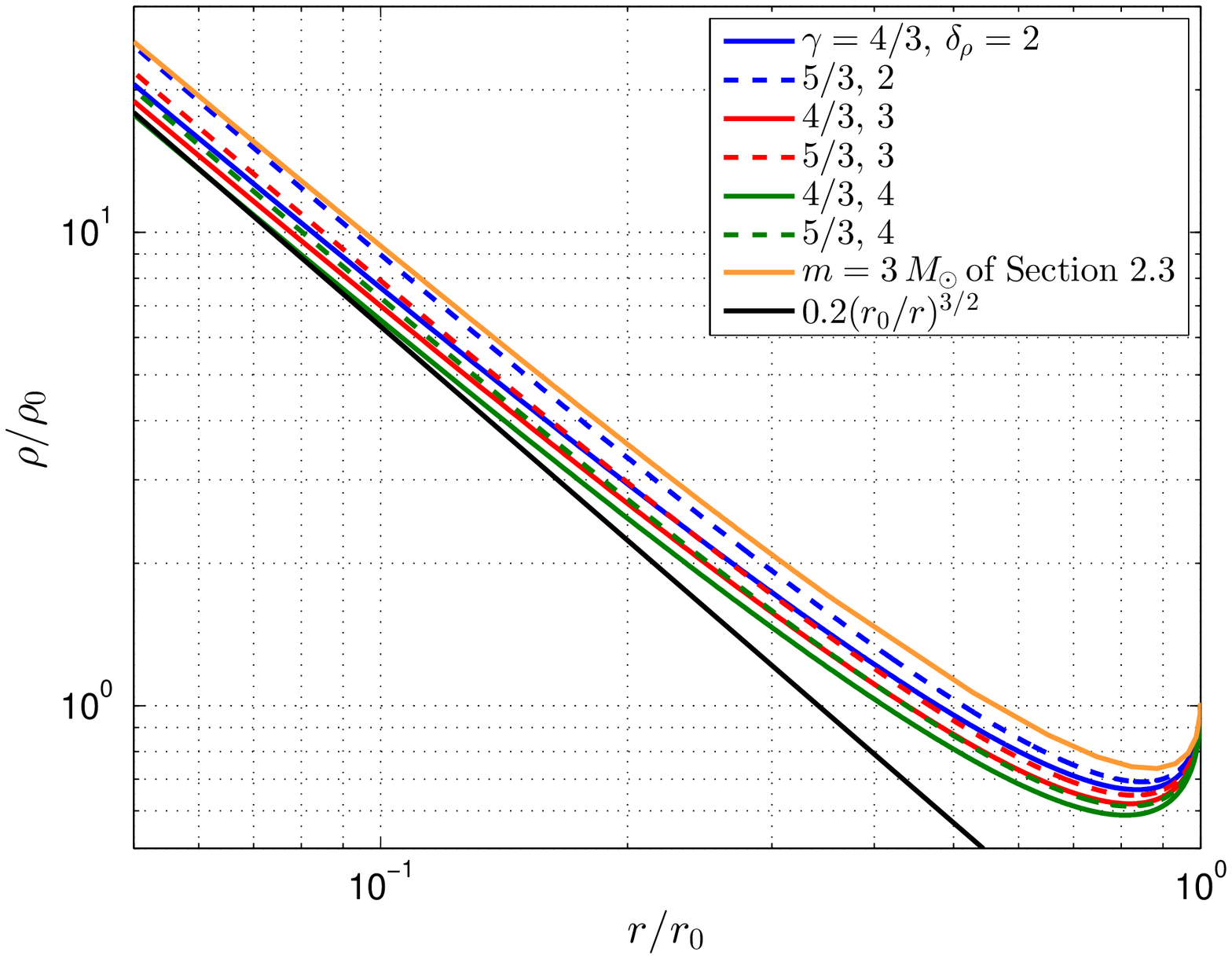}
 }
 \subfigure[]{
             \includegraphics[width=0.5\textwidth]{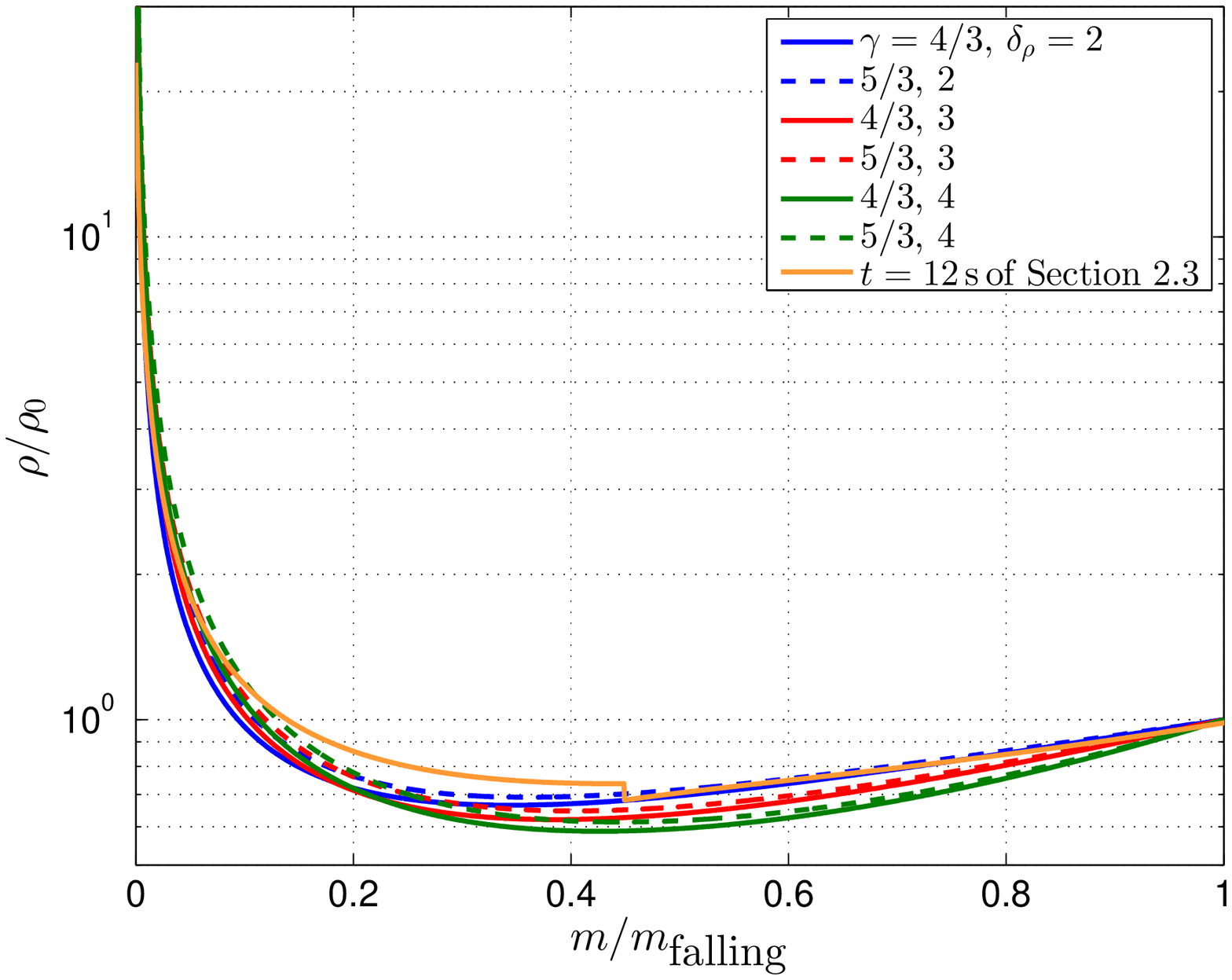}
 }
      \caption{Self-similar collapse. Panel (a) shows the compression of a Lagrangian element as a function of its radius (normalized to its original radius) for different values of the adiabatic index $\gamma$ and the density power-law index $\der$. For comparison, the compression of the $m=3\,M_{\odot}$ mass element from the simulation of Section~\ref{sec:example} is shown. Panel (b) shows the compression as a function of mass at a given time, normalized to the total mass of falling material. For comparison, the compression at $t=12\,\textrm{s}$ from the simulation of Section~\ref{sec:example} is shown. \label{fig:SeS}}
\end{figure}

In order to study the collapse of a mass element, we make the following approximations about the profile in its neighborhood: (1) $\rho$ is a power-law in radius $\rho\propto r^{-\delta_{\rho}}$ , (2) the accumulated mass is independent of radius, and (3) adiabatic compression is described by a constant adiabatic index $\gamma$. Under these reasonable approximations, the flow is described by a self-similar solution that is found in the appendix. Note that while the self-similar solution assumes these assumptions to hold throughout the profile, the evolution of any given mass element is not sensitive to the profile at distant radii and thus the results are approximately correct for general profiles. The compression of a mass element as a function of its radius is shown in panel (a) of Figure~\ref{fig:SeS} for various values of the power-law index $\delta_{\rho}$ and the adiabatic index $\gamma$. As can be seen, as the radius decreases, the density first decreases and then increases, approaching a compression of $\mysim 10$ at $r/r_0=0.1$. For comparison, the compression of the $m=3\,M_{\odot}$ mass element from the simulation of Section~\ref{sec:example} is shown. The composition of this mass element is pure oxygen and negligible burning occurs during the compression. As can be seen in the figure, the self-similar solutions agree with the numerical compression to an accuracy of $\mysim10\%$.

The compression at small radii can be obtained as follows. Consider two adjacent mass elements that start at $r_0$ and are initially separated by $dr_0$.
The rarefaction wave reaches the two elements at slightly different times separated by $dt_0$. As the rarefaction wave moves at the speed of sound $c_{\textrm{s}0}$ we have (using Equation \eqref{eq:SeS_c})
\begin{equation}\label{eq:Simple_dr_0}
dr_0=c_{\textrm{s}0}dt_0=\sqrt{\frac{\gamma GM}{(\der+1) r_0}}dt_0.
\end{equation}
As the elements fall they reach each radius $r$ at slightly different times separated by $dt=dr/v$, where $dr$ is their instantaneous separation and $v$ is their velocity. At small radii, the elements approach free-fall and therefore $v=(2GM/r)^{1/2}$ so that
\begin{equation}\label{eq:Simple_dr}
dr=vdt=\sqrt{\frac{2GM}{r}}dt.
\end{equation}
The asymptotic compression is thus given by
\begin{equation}
\frac{\rho}{\rho_0}=\frac{r_0^2dr_0}{r^2dr}=\sqrt{\frac{\gamma}{2(\der+1)}}\frac{dt_0}{dt}(r/r_0)^{-3/2}.
\end{equation}
At small radii, $dt$ approaches a constant $dt_f$--the time difference between the arrival at $r=0$, and we can approximate
\begin{equation}
\frac{\rho}{\rho_0}=\sqrt{\frac{\gamma}{2(\der+1)}}\left(\frac{dt_f}{dt_0}\right)^{-1}(r/r_0)^{-3/2}.
\end{equation}
The values of $dt_f/dt_0$ for different choices of $\der$ and $\gamma$ are given in Table~\ref{tbl:sc}.
As can be seen, $dt_f/dt_0\approx 2$. We therefore expect that at small $r$,
\begin{equation}\label{eq:Simple_AssymptoticRho}
\frac{\rho}{\rho_0}\approx 0.2 (r/r_0)^{-3/2},
\end{equation}
which is consistent with the results shown in panel (a) of Figure~\ref{fig:SeS}.

The amount of time spent at small radii is short and thus the mass at any given time that is significantly compressed is small, as seen in panel (b) of Figure~\ref{fig:SeS}, which shows the compression as a function of mass at a given time, normalized to the total mass of falling material (i.e., with negative velocity), $m_{\textrm{falling}}$. For comparison, the compression at $t=12\,\textrm{s}$ from the simulation of Section~\ref{sec:example} is shown.  As can be seen in the figure, the self-similar solutions agree with the numerical compression to an accuracy of $\mysim10\%$. Note that the small density jump at $m/m_{\textrm{falling}}\approx 0.45$ is caused by a small amount of thermonuclear burning that is not present in the self-similar solutions.

As the density of a falling mass element becomes higher, the temperature rises due to the adiabatic compression. In Section~\ref{sec:Explosion} we show that the pre-collapse electron-to-photon number ratio is of order unity. We next show that under such conditions the photon-to-electron density remains practically constant during the adiabatic compression, allowing the temperature to be easily calculated. The equation of state is approximated by
\begin{equation}\label{eq:Simple_EOS}
p=n_{\textrm{e}}T+a_{\textrm{R}}T^4/3=(n_{\textrm{e}}+n_{\gamma})T,
\end{equation}
where
\begin{equation}\label{eq:Simple_ne}
n_{\textrm{e}}=\frac{\rho}{2m_p}, ~n_{\gamma}=\frac{a_{\textrm{R}}T^{3}}{3},
\end{equation}
are the electron and photon densities, respectively, and $a_{\textrm{R}}$ is the blackbody radiation constant.

During adiabatic compression we have
\begin{equation}\label{eq:adiabatic}
d\left (\frac{e}{n_{\textrm{e}}}\right)=-pd\left(\frac{1}{n_{\textrm{e}}}\right)=\frac{p}{n_{\textrm{e}}}\frac{dn_{\textrm{e}}}{n_{\textrm{e}}},
\end{equation}
where $e$ is the energy density and is given by
\begin{equation}\label{eq:Simple_EOS_e}
e=\frac32n_{\textrm{e}}T+a_{\textrm{R}}T^4=3\left(\frac12 n_{\textrm{e}}+n_{\gamma}\right)T.
\end{equation}

Using the fact that $(n_{\gamma}/n_{\textrm{e}})n_{\textrm{e}}=a_{\textrm{R}}T^3/3$, we find
\begin{equation}\label{eq:dT_T}
3\frac{dT}T=\frac{dx_{\gamma}}{x_{\gamma}}+\frac{dn_{\textrm{e}}}{n_{\textrm{e}}},~~~~~\textrm{where}~~x_{\gamma}\equiv\frac{n_{\gamma}}{n_{\textrm{e}}}.
\end{equation}
Equations \eqref{eq:Simple_EOS}, and \eqref{eq:adiabatic}--\eqref{eq:dT_T} can be used to obtain
\begin{equation}
\frac{dn_{\textrm{e}}}{n_{\textrm{e}}}=\frac{dx_{\gamma}}{x_{\gamma}}+8dx_{\gamma},
\end{equation}
implying that during adiabatic compression we have
\begin{equation}\label{eq:Simple_adiabatic}
\frac{n_{\gamma}}{n_{\textrm{e}}}\exp\left(8\frac{n_{\gamma}}{n_{\textrm{e}}}\right)\propto n_{\textrm{e}}\propto \rho.
\end{equation}
In order to change the photon-to-electron ratio from $0.5$ to $1$ ($1$ to $2$), $n_{\textrm{e}}$ needs to be compressed by a factor of about $110$ ($6000$), implying that for $n_{\gamma}/n_{\textrm{e}}\gtrsim0.5$, it is approximately constant for a very wide range of compressions. The temperature follows $T\propto n_{\gamma}^{1/3}\propto (n_{\textrm{e}}/n_{\gamma})^{1/3}n_{\textrm{e}}^{1/3}$, and is therefore proportional to $\rho^{1/3}$ to an excellent approximation. Once the element reaches radii much smaller than its initial radius, we can use Equation \eqref{eq:Simple_AssymptoticRho} to find
\begin{equation}\label{eq:AssimptoticT}
T\approx 0.6 (r/r_0)^{-1/2}T_0.
\end{equation}
At $r\sim 0.1r_0$ the temperature rises by about a factor of $2$, typically allowing for a much higher burning rate. At sufficiently fast burning rates, significant energy can be released on a timescale that is shorter than the sound crossing time and the free-fall time and a detonation wave forms.

Once a detonation wave forms, its velocity with respect to the local rest-frame must be larger than the infall velocity so that it propagates out. Assuming that the energy release is much larger than the thermal energy, the shock velocity is given by the Chapman--Jouguet velocity
\begin{equation}
v_{s}=\sqrt{2Q(\gamma^2-1)}\approx1.2\cdot 10^9\left(\frac{Q}{\textrm{ MeV}/b}\right)^{1/2}\textrm{cm}\,\textrm{s}^{-1},
\end{equation}
where the appropriate $\gamma\approx 4/3$ was used.
In order for $v_{s}>v_{\textrm{ff}}=(2GM/r)^{1/2}$, the detonation must be ignited at a sufficiently large radius
\begin{equation}
r_{\textrm{det}}>\frac{GM}{Q(\gamma^2-1)}\approx 5\cdot 10^{8}\left(\frac{Q}{{\textrm{MeV}}/b}\right)^{-1}\frac{M}{3M_{\odot}}\,\textrm{cm}.
\end{equation}
Given that compression requires a change in radius of at least $2$, the material that can ignite the detonation must initially be at a radius greater than about $10^9\,\textrm{cm}$. This constraint is satisfied in the successful numerical explosions described in Section~\ref{sec:simulation}.

\subsection{Explosion}\label{sec:Explosion}
We next derive the constraints on the initial profile required so that a considerable mass is traversed by the detonation wave before it fails.  The detonation wave requires high densities to propagate so that the released thermonuclear energy (per unit volume) $\mysim Q\rho$ is sufficiently high to increase the temperature to values  of $T>T_c\sim 10^9\,\textrm{K}$, where the burning is faster than the free-fall time. The threshold density $\rho_{\textrm{det}}$ is roughly given by
\begin{equation}
\rho_{\textrm{det}} \mysim a_{\textrm{R}}T_c^4/Q\approx 10^4 \left(\frac{T_c}{10^9\,\textrm{K}}\right)^4\left(\frac{Q}{\textrm{MeV}/b}\right)^{-1}\,\textrm{g}\,\textrm{cm}^{-3}.
\end{equation}
As explained above (panel (b) of Figure~\ref{fig:SeS}), the amount of material that is significantly compressed at any given time is very small. The initial profile must therefore contain $\mygtrsim M_{\odot}$ of explosive material at high densities $\rho\gtrsim\rho_{\textrm{det}}$. In addition, the mass available for the explosion must initially have a low temperature $T<T_c$ to avoid fast burning prior to the collapse. As we next show, the amount of mass with a high density and low temperature is tightly constrained by the requirement of a hydrostatic equilibrium. This is the basic reason for the fine tuning required in Section~\ref{sec:sensitivity}.

For simplicity consider a density profile with an inner core mass $M_{\textrm{in}}$ within $r_{\textrm{in}}=2\cdot 10^8\,\textrm{cm}$ and a uniform mass per logarithmic radius interval at larger radii,
\begin{equation}\label{eq:Mlogdef}
\rho=\frac{\Mlog}{4\pi r^3}.
\end{equation}
The enclosed mass within a radius $r$ is
\begin{equation}
M(r)=M_{\rm in}+\Mlog\ln(r/r_{\rm in})
\end{equation}
and assuming hydrostatic equilibrium the pressure is given by
\begin{equation}\label{eq:Simple_p_exact}
p=\int_r^{\infty}\frac{GM(r')\rho(r')}{r'^2}dr'=\frac{G(M+\Mlog/4)\Mlog}{16\pi r^4}.
\end{equation}

\begin{figure}
\subfigure[]{
             \includegraphics[width=0.5\textwidth]{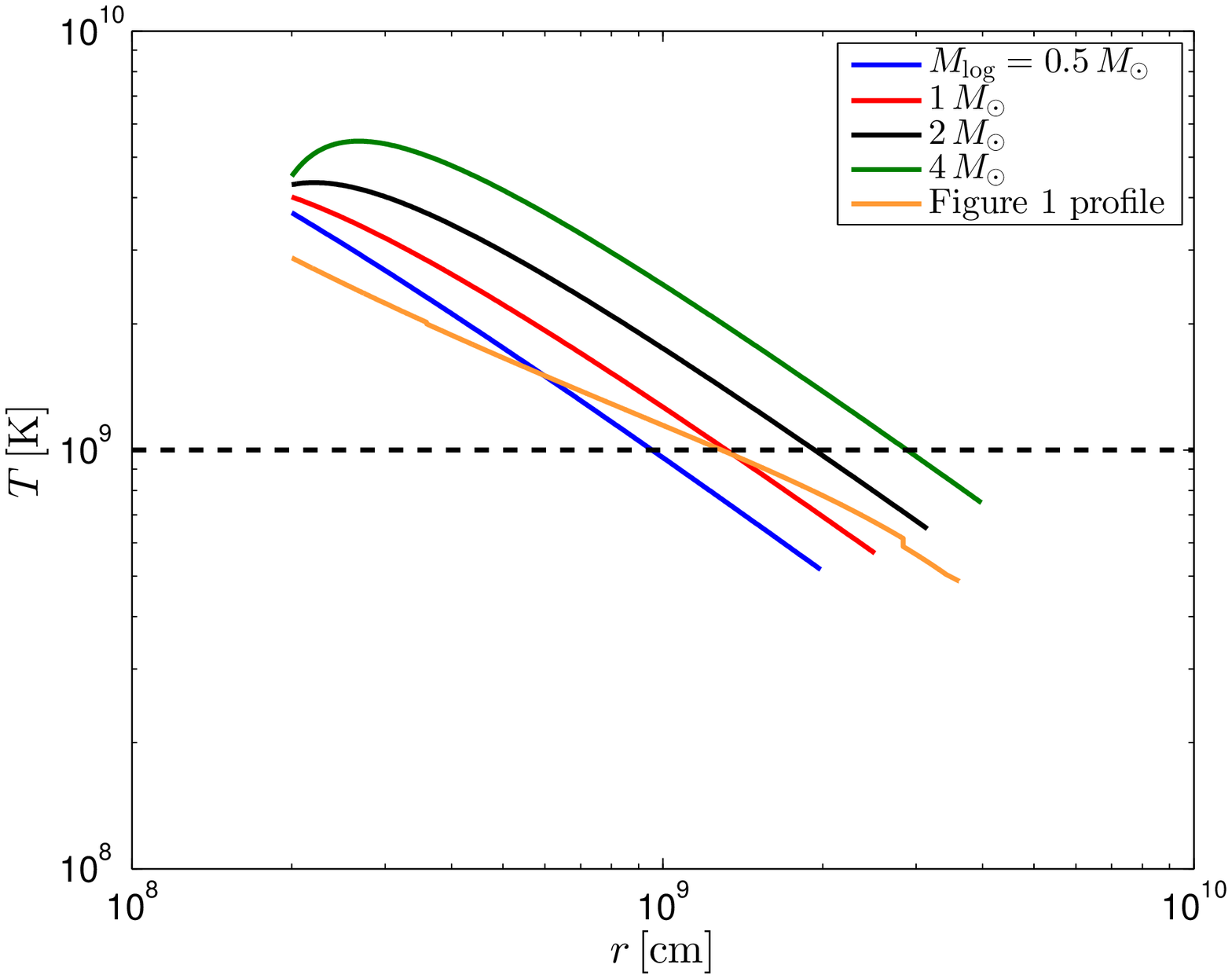}
 }
 \subfigure[]{
             \includegraphics[width=0.5\textwidth]{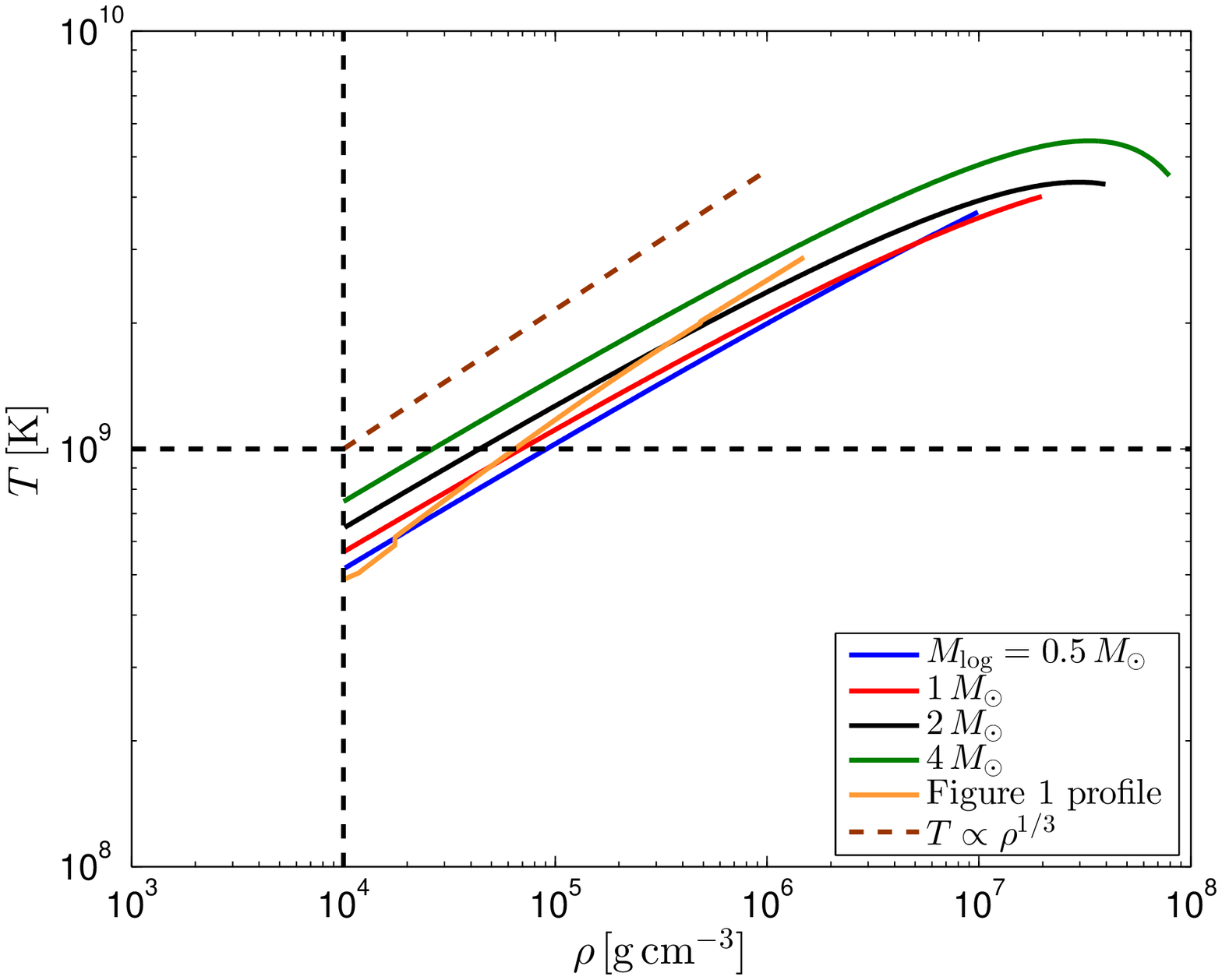}
 }
      \caption{Simple model. For $\Mlog/M_{\odot}=0.5,1,2,4$ the mass that satisfies the requirements $T<10^9\,\textrm{K}$ and $\rho>10^4\,\textrm{g}\,\textrm{cm}^{-3}$ is found to be approximately $0.4,0.7, 1,1.3~M_{\odot}$, respectively. The profile from Figure~\ref{fig:InitialProfile} is shown for comparison, with $\myapprox1.9\,M_{\odot}$, which satisfies the above requirements. \label{fig:Simple_T_r_rho}}
\end{figure}

Using these approximations, the temperature can be readily found at each radius, given $M_{\textrm{in}}$ and $\Mlog$. As in Section~\ref{sec:1D simulations}, we adopt an inner mass of $M_{\textrm{in}}=1.2\,M_{\odot}$. The temperature and density profiles for a range of density normalizations, $\Mlog$, are shown in Figure~\ref{fig:Simple_T_r_rho} at radii where $\rho>10^4\,\textrm{g}\,\textrm{cm}^{-3}$. As can be seen in the figure, the range of radii where there are sufficiently low temperatures $T<10^9\,\textrm{K}$ and sufficiently high densities $\rho>10^4\,\textrm{g}\,\textrm{cm}^{-3}$ is narrow and smaller for higher density normalizations. The amount of mass that satisfies this constraint reaches about $1.3\,M_{\odot}$ for the high density normalization  $\Mlog=4\,M_{\odot}$, significantly limiting the amount of available thermonuclear energy, and is lower for smaller $\Mlog$. For comparison, the profile from Figure~\ref{fig:InitialProfile} is also shown. As can be seen, while the profile is shallower (mostly due to the deviation from $\rho\propto r^{-3}$), the simple model is confined between the normalization range of $\Mlog= 0.5$--$3\,M_{\odot}$ in the range $5\cdot 10^8\,\textrm{cm} < r< 4\cdot 10^9\,\textrm{cm}$. This explains the narrow range of parameters that allows successful explosions, which was obtained in Section~\ref{sec:sensitivity}. In order to understand the origin of this tight constraint, the temperature is plotted as a function of the density in panel (b) of Figure~\ref{fig:Simple_T_r_rho}. As can be seen the temperatures and densities are related by $T\propto \rho^{1/3}$ with little dependence on $\Mlog$ (also holds for the profile from Figure~\ref{fig:InitialProfile}). This reflects the fact that for massive extended stars in hydrostatic equilibrium, the ratio of the photon density $n_{\gamma}=a_{\textrm{R}}T^3/3$ to the electron number density $n_{\textrm{e}}\approx \rho/(2m_p)$ is of order unity, as we next demonstrate.

To a good approximation, Equation~\eqref{eq:Simple_p_exact} can be written as
\begin{equation}\label{eq:Simple_p_approx}
p=\frac{GM\rho}{4r}.
\end{equation}
Using Equations~\eqref{eq:Simple_p_approx},~\eqref{eq:Mlogdef}, and~\eqref{eq:Simple_ne}, we find
\begin{equation}\label{eq:Simple_ne4p3}
\frac{p^3}{n_{\textrm{e}}^4}=\frac{\pi G^3M^3m_p^4}{\Mlog}.
\end{equation}
Using the equation of state, \eqref{eq:Simple_EOS}, we find
\begin{equation}\label{eq:ne4p3_EOS}
\frac{p^3}{n_{\textrm{e}}^4}=\frac{3}{a_{\textrm{R}}}\frac{n_{\gamma}}{n_{\textrm{e}}}\left(1+\frac{n_{\gamma}}{n_{\textrm{e}}}\right)^3.
\end{equation}

Equating Equations~\eqref{eq:ne4p3_EOS} and \eqref{eq:Simple_ne4p3} we find
\begin{equation}\label{eq:Simple_nga_ne}
\frac{n_{\gamma}}{n_{\textrm{e}}}\left(1+\frac{n_{\gamma}}{n_{\textrm{e}}}\right)^3\approx0.41\frac{M^3}{\Mlog M_{
\rm ch}^2}
\end{equation}
where
\begin{equation}
M_{\rm ch}\approx3.098\left(\frac{\hbar c}{G}\right)^{3/2}\mu_e^{-2}\approx2.51(a_{\textrm{R}}G^3)^{-1/2}\mu_e^{-2}
\end{equation}
is the Chandrasekhar mass and $\mu_e=2m_p$.

Since $M\gtrsim 2\Mlog,M_{\rm ch}$, the right hand side is larger than unity. For $2<M^3/(\Mlog M_{\rm ch}^2)<100$, we have $0.7<(n_{\gamma}/n_{\textrm{e}})^{1/3}<1.22$.

The temperature can be expressed as
\begin{align}\label{eq:Simple_Trho}
T&=\left(\frac{3\rho}{2 m_p a_{\textrm{R}}}\right)^{1/3}\left(\frac{n_{\gamma}}{n_{\textrm{e}}}\right)^{1/3}\cr
&\approx5.5\cdot 10^8\left(\frac{\rho}{10^4\,\textrm{g}\,\textrm{cm}^{-3}}\right)^{1/3}\left(\frac{n_{\gamma}}{n_{\textrm{e}}}\right)^{1/3}\,\textrm{K}\cr
\end{align}
and for $(n_{\gamma}/n_{\textrm{e}})^{1/3}\approx 1$, agrees with the values presented in panel (b) of Figure~\ref{fig:Simple_T_r_rho}.

Equation \eqref{eq:Simple_Trho} demonstrates that there is a tight region for which the density can be high $\rho\gtrsim 10^4\,\textrm{g}\,\textrm{cm}^{-3}$ while the temperature is low $T\lesssim 10^9\,\rm K$. Moreover, by relating the temperature to the radius,
\begin{equation}\label{eq:Simple_Tr}
T=1.4\cdot 10^9\left(\frac{r}{10^9\,\textrm{cm}}\right)^{-1}\left(\frac{\Mlog}{M_{\odot}}\right)^{1/3}\left(\frac{n_{\gamma}}{n_{\textrm{e}}}\right)^{1/3}\,\textrm{K},
\end{equation}
we see that the small range in temperature corresponds to a small range in radii and therefore a limited amount of mass available for thermonuclear burning by detonation.

\section{Discussion}
\label{sec:discussion}

In this paper we revisited the collapse-induced thermonuclear supernovae mechanism. In Section~\ref{sec:1D simulations} we performed a series of 1D calculations of collapsing massive stars with simplified initial density profiles and various compositions, assuming that the neutrinos had a negligible effect on the outer layers. We demonstrate that $\mysim10\,\textrm{s}$ after the core-collapse of a massive star, a successful thermonuclear explosion of the outer shells is possible for some initial density and composition profiles that include a significant layer of He--O mixture.

There are several challenges in associating these simulations with observed supernovae.
\begin{enumerate}
\item Post-collapse synthesized material, and in particular $^{56}$Ni is not released in the simulations.
\item The obtained kinetic energies of the ejecta are limited to $\mylesssim 10^{50}\,\textrm{erg}$, which is not sufficient for explaining typical observed type II supernovae.
\item The required profiles are tuned (Section~\ref{sec:sensitivity}) and require the presence of a mixture of He and O with burning times of $\mylesssim100\,\textrm{s}$ ($\myapprox10$ times the free-fall time) prior to collapse, which is not currently expected in stellar evolution models.
\end{enumerate}

In Section~\ref{sec:simple} we used simple arguments to demonstrate that for a general family of profiles, satisfying some reasonable constrains, strong explosions may only be possible for a narrow range of density amplitudes. The detonation wave requires high densities of $\mygtrsim 10^4\,\textrm{g}\,\textrm{cm}^{-3}$ to propagate. While the elements are compressed adiabatically as they fall, only a small mass is significantly compressed at any given time (see panel (b) of Figure~\ref{fig:SeS}). A successful explosion thus requires a significant mass of explosive material $M\gtrsim M_{\odot}$ to be present in the initial profile at high densities. The high required densities are contrasted by the requirement for low initial temperatures $T\lesssim 10^9\,\textrm{K}$ so that the pre-collapse burning rate is much slower than the free-fall time. Indeed, hydrostatic equilibrium requires a roughly equal number of photons and electrons where significant mass is present (see Equation~\eqref{eq:Simple_nga_ne}), implying that high densities $\rho\gtrsim 10^4\,\textrm{g}\,\textrm{cm}^{-3}$ require high temperatures $T\gtrsim 5\cdot 10^8\,\textrm{K}$ (see Equation~\eqref{eq:Simple_Trho}).

While the 1D collapse scenarios studied here are therefore unlikely to represent the majority of observed type II supernovae, they serve as a proof of concept that core-collapse-induced thermonuclear explosions are possible. In fact, as far as we know, these are the first set of 1D simulations, based on first-principles physics, where a supernova is convincingly demonstrated to occur following core-collapse. The crucial ingredient in this scenario is the ignition of a detonation wave, which is fully resolved here for the first time (Section~\ref{sec:ignition}). Further studies are required to examine whether more realistic simulations (in particular multi-dimensional) may lead to explosions that better agree with observations and stellar evolution constraints. Unlike neutrino driven explosions, which require the solution of nonthermal transport equations, 3D simulations of this thermonuclear mechanism are possible with current computational capabilities.

\begin{figure}
\includegraphics[width=0.5\textwidth]{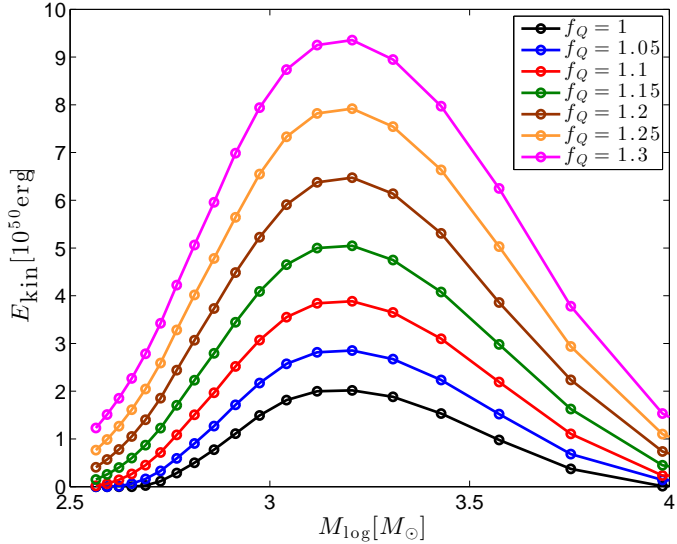}
\caption{Asymptotic kinetic energy of the ejecta as a function of $\Mlog$ for $\Mcore=10\,M_{\odot}$, $r_{\textrm{O}/\textrm{He}}=1$ and $t_{\textrm{b},0}/t_{\textrm{ff},0}=10$, with artificially increased available thermonuclear energy per unit mass. The legend indicates $f_{Q}$, the increase difference in binding energy between the initial composition and the final composition (assumed to be pure silicon), achieved by artificially changing the binding energy of helium.
\label{fig:fQ}}
\end{figure}

An interesting property of the core-collapse-induced thermonuclear explosions reported here is the fact that the potential energy of the star canceled  most of the released thermonuclear energy. This means that even a small increase in the released thermonuclear energy can significantly increase the obtained kinetic energy of the ejecta. To demonstrate this, we rerun the set of simulations with $\Mcore=10\,M_{\odot}$, $r_{\textrm{O}/\textrm{He}}=1$ and $t_{\textrm{b},0}/t_{\textrm{ff},0}=10$, with increased available thermonuclear energy per unit mass. To achieve this, we change the binding energy of helium, such that the difference in binding energy between the initial composition and the final composition (assumed to be pure silicon) increased by a factor $f_{Q}$. The results shown in Figure~\ref{fig:fQ} indicate that $f_{Q}=1.3$ is enough to increase the kinetic energy of the ejecta to $\mysim10^{51}\,\textrm{erg}$ (which includes post-collapse synthesized material).

One physical process that cannot be treated in 1D and that may play an important role is rotation. In fact, preliminary 2D calculations that include rotation (not reported here) indicate that stronger explosions are possible for a wider range of initial conditions \citep[][]{Kushnir2015}. In addition, post-collapse synthesized material is ejected. This is different from the results of a previous study that included rotation where an ignition of a detonation wave was not obtained \citep[][]{Bodenheimer}, with the main difference likely being the presence of He--O mixtures in \citet{Kushnir2015}.

\acknowledgments We thank R. Waldman and E. Livne for useful discussions. D.~K. gratefully acknowledges support from Martin A. and Helen Chooljian Founders' Circle and from the Friends of the Institute for Advanced Study. FLASH was in part developed by the DOE NNSA-ASC OASCR Flash Center at the University of Chicago. Computations were performed at IAS cluster.


\bibliographystyle{apj}

\begin{appendix}
\section{Self-similar collapse}\label{sec:SeS}

Consider an initial profile with a power-law density distribution
\begin{equation}
\rho_0\propto r_0^{-\der},
\end{equation}
where $r_0$ is the initial radius of each mass element. For simplicity, the enclosed mass is assumed to be independent of radius and denoted by $M$ (when applying the results to analyze the behavior of a mass element at $r$ in actual profiles, $M$ should be substituted with the enclosed mass $M(r)$). The initial  pressure profile can be calculated and is given by,
\begin{equation}
p_0=\int_{r_0}^{\infty}\rho_0\frac{GM}{r_0^2}dr_0=\frac{1}{\der+1}\frac{GM}{r_0}\rho_0,
\end{equation}
while the speed of sound is,
\begin{equation}\label{eq:SeS_c}
c_{\textrm{s}0}=\sqrt{\frac{\gamma p_0}{\rho_0}}=\sqrt{\frac{\gamma}{\der+1}}\sqrt{\frac{GM}{r_0}},
\end{equation}
where $\gamma\approx 4/3$ is the adiabatic index. The refraction wave moves at the speed of sound and reaches a given mass element $r_0$ at a time
\begin{equation}\label{eq:t0}
t_0=\int_0^{r_0} \frac{dr_0}{c_{\textrm{s}0}}=\frac23\frac{r_0}{c_{\textrm{s}0}}=\frac23\sqrt{\frac{\der+1}{\gamma GM}}r_0^{3/2}.
\end{equation}

Since there is no scale in this problem, the hydrodynamic collapse is self-similar and the position as a function of time can be given as
\begin{equation}\label{eq:SES}
r(r_0,t)=r_0R(s)~~~~~\rho(r_0,t)=\frac{\rho_0}{Q(s)}~~~~~p=p_0Q^{-\gamma}
\end{equation}
where
\begin{equation}
s=\frac{t}{t_0}\propto t r_0^{-3/2},
\end{equation}
and $R(s)$ and $Q(s)$ are functions that depend on $s$ alone (note that in this section $Q$ has nothing to do with the available thermonuclear energy). By employing conservation of mass and momentum, we next derive two first order ordinary differential equations of the form $R'=f(R,Q,s)$ and $Q'=g(R,Q,s)$, where prime denotes a derivative with respect to $s$. These equations can be easily integrated numerically to obtain $R$ and $Q$ and therefore the entire collapsing flow.

Mass conservation implies that
\begin{equation}
\rho r^2dr=\rho_0r_0^2dr_0.
\end{equation}
To relate $dr$ to $dr_0$ note that
\begin{equation}
\frac{ds}{dr_0}=-\frac32\frac{s}{r_0},
\end{equation}
and
\begin{equation}\label{eq:dr_dr0}
\frac{dr}{dr_0}=R-\frac32sR',
\end{equation}
where derivatives with respect to $r$ or $r_0$ are taken at a constant time $t$.
Using equations \eqref{eq:SES} and \eqref{eq:dr_dr0} we have
\begin{equation}\label{eq:Q_of_R}
Q=\frac{r^2dr}{r_0^2dr_0}=R^2(R-\frac32sR'),
\end{equation}
from which we obtain an equation for $R'$,
\begin{equation}\label{eq:QR1}
R'=\frac23\frac{R^3-Q}{sR^2}.
\end{equation}

Momentum conservation implies that
\begin{equation}\label{eq:force}
\ddot r=-\frac{1}{\rho}\frac{dp}{dr}-\frac{GM}{r^2}.
\end{equation}
It is straightforward to find the following expressions for each of the three terms in Equation~\eqref{eq:force},
\begin{equation}\label{eq:ddotr}
\ddot r=\frac{r_0}{t^2_0}R''=\frac94\frac{\gamma}{\der+1}\frac{GM}{r_0^2}R'',
\end{equation}
\begin{equation}\label{eq:dpdr_rho}
\frac{1}{\rho}\frac{dp}{dr}=\frac{GM}{r_0^2}\frac{R^2}{\der+1}\left[\frac32 s\gamma Q^{-(\gamma+1)}Q'-(\der+1)Q^{-\gamma}\right],
\end{equation}
and
\begin{equation}\label{eq:GM_r2}
\frac{GM}{r^2}=\frac{1}{R^2}\frac{GM}{r_0^2},
\end{equation}
resulting in
\begin{equation}\label{eq:Rpp}
\frac94\frac{\gamma}{\der+1}R''=R^2Q^{-\gamma}-\frac{1}{R^2}-\frac32 s\gamma\frac{R^2}{\der+1} Q^{-(\gamma+1)}Q'.
\end{equation}
By differentiating Equation~\eqref{eq:Q_of_R} with respect to $s$, we get
\begin{equation}\label{eq:Qp}
Q'=2RR'(R-\frac32sR')+R^2(R'-\frac32R'-\frac32sR'').
\end{equation}

Equations~\eqref{eq:Rpp} and~\eqref{eq:Qp} involve $R,R',R'',Q,Q'$ and $s$ and can be used to express $Q'$ in terms of $Q,R,R'$ and $s$,
\begin{equation}\label{eq:QR2}
Q'=\frac{\frac{R'}{s}\left(\frac{2Q}{R}-\frac{R^2}{2}\right)+\frac23\frac{\der+1}{\gamma}(1-R^4Q^{-\gamma})}{(\frac1s-sR^4Q^{-(\gamma+1)})}.
\end{equation}

At $s=1$ we have $R(1)=Q(1)=1$ by construction. Equations \eqref{eq:QR1} and \eqref{eq:QR2} can be integrated from $s=1$ up to a point $s_c$ where $R(s_c)=0$. Note that at $s=1$, the denominator in Equation~\eqref{eq:QR2} vanishes and we need to start the integration from some value $s$ close to $1$ with appropriate asymptotic conditions. The self-consistency of these equations implies that for $s$ close to unity, $s=1+ds$, we have
\begin{align}
R(1+ds)&= 1+O(ds^2)\cr
Q(1+ds)&=1+\frac{2+\frac23(\der+1)}{\gamma+1}ds+O(ds^2).\cr
\end{align}
The obtained values of $s_c$ are given in Table~\ref{tbl:sc}. $s_c$ can be used to find the time it takes for a mass element that started at $r_0$ to get to zero,
\begin{equation}\label{eq:tf}
t_f=s_ct_0,
\end{equation}
from which we have
\begin{equation}\label{eq:dtfdt0}
\frac{dt_f}{dt_0}=s_c.
\end{equation}

At a given time $t$, the original location of each element can be related to $s$ by $t/t_0=s$, implying that
\begin{equation}
r_0\propto s^{-2/3}
\end{equation}
and
\begin{equation}\label{eq:SeS_dmds}
dm\propto r_0^2dr_0\rho_0\propto d(r_0^{3-\der})\propto  d(s^{-2+2\der/3})
\end{equation}
where for $\der=3$ we have $dm\propto d\log(s)$ instead.
At small values of $r$, where $s\approx s_c\approx 2$, we have $dm\propto (s_c-s)$. The compression factor is $\rho/\rho_0\propto (r/r_{0})^{-3/2}\propto (s_c-s)^{-1}$, explaining the fact that high compression is only possible for a small amount of mass, as seen in Figure~\ref{fig:SeS}.

\end{appendix}
\begin{deluxetable}{ccccccc}
\tablecaption{The simulations in which the asymptotic kinetic energy of the ejecta is larger than $5\cdot10^{49}\,\textrm{erg}$ \label{tbl:list}}
\tablewidth{0pt} \tablehead{ \colhead{$\Mcore\,[M_{\odot}]$} & \colhead{$\rho_{in}\,[10^{6}\,\textrm{g}\,\textrm{cm}^{-3}]$} & \colhead{$\Mlog\,[M_{\odot}]$} & \colhead{$r_{\textrm{O}/\textrm{He}}$} & \colhead{$t_{\textrm{b},0}/t_{\textrm{ff},0}$} & \colhead{$r_{\textrm{base}}\,[10^{9}\,\textrm{cm}]$} & \colhead{$E_{\textrm{kin}}\,[10^{50}\,\textrm{erg}]$}}
\startdata
10	&	0.9	&	3.57	&	1	&	2	&	1.81	&	0.72	\\
10	&	1.0	&	3.43	&	1	&	2	&	1.96	&	1.4	\\
10	&	1.1	&	3.31	&	1	&	2	&	2.09	&	2.1	\\
10	&	1.2	&	3.21	&	1	&	2	&	2.20	&	2.5	\\
10	&	1.3	&	3.12	&	1	&	2	&	2.28	&	2.6	\\
10	&	1.4	&	3.04	&	1	&	2	&	2.35	&	2.7	\\
10	&	1.5	&	2.97	&	1	&	2	&	2.40	&	2.5	\\
10	&	1.6	&	2.91	&	1	&	2	&	2.43	&	2.3	\\
10	&	1.7	&	2.86	&	1	&	2	&	2.46	&	1.9	\\
10	&	1.8	&	2.81	&	1	&	2	&	2.48	&	1.5	\\
10	&	1.9	&	2.77	&	1	&	2	&	2.49	&	1.1	\\
10	&	2.0	&	2.73	&	1	&	2	&	2.49	&	0.78	\\
10	&	0.9	&	3.57	&	1	&	10	&	2.25	&	0.98	\\
10	&	1.0	&	3.43	&	1	&	10	&	2.41	&	1.5	\\
10	&	1.1	&	3.31	&	1	&	10	&	2.53	&	1.9	\\
10	&	1.2	&	3.21	&	1	&	10	&	2.62	&	2.0	\\
10	&	1.3	&	3.12	&	1	&	10	&	2.69	&	2.0	\\
10	&	1.4	&	3.04	&	1	&	10	&	2.74	&	1.8	\\
10	&	1.5	&	2.97	&	1	&	10	&	2.77	&	1.5	\\
10	&	1.6	&	2.91	&	1	&	10	&	2.79	&	1.1	\\
10	&	1.7	&	2.86	&	1	&	10	&	2.80	&	0.77	\\
10	&	1.1	&	3.28	&	3/2	&	10	&	2.52	&	0.61	\\
10	&	1.2	&	3.18	&	3/2	&	10	&	2.61	&	0.65	\\
10	&	1.3	&	3.10	&	3/2	&	10	&	2.67	&	0.61	\\
10	&	1.4	&	3.02	&	3/2	&	10	&	2.72	&	0.51	\\
10	&	1.0	&	3.46	&	2/3	&	10	&	2.39	&	0.57	\\
10	&	1.1	&	3.34	&	2/3	&	10	&	2.52	&	0.74	\\
10	&	1.2	&	3.23	&	2/3	&	10	&	2.62	&	0.79	\\
10	&	1.3	&	3.14	&	2/3	&	10	&	2.69	&	0.73	\\
10	&	1.4	&	3.06	&	2/3	&	10	&	2.75	&	0.62	\\
8	&	1.0	&	2.61	&	1	&	10	&	1.99	&	0.68	\\
8	&	1.1	&	2.51	&	1	&	10	&	2.11	&	1.1	\\
8	&	1.2	&	2.43	&	1	&	10	&	2.20	&	1.4	\\
8	&	1.3	&	2.36	&	1	&	10	&	2.27	&	1.6	\\
8	&	1.4	&	2.30	&	1	&	10	&	2.32	&	1.7	\\
8	&	1.5	&	2.25	&	1	&	10	&	2.36	&	1.7	\\
8	&	1.6	&	2.20	&	1	&	10	&	2.39	&	1.6	\\
8	&	1.7	&	2.16	&	1	&	10	&	2.41	&	1.4	\\
8	&	1.8	&	2.12	&	1	&	10	&	2.43	&	1.2	\\
8	&	1.9	&	2.09	&	1	&	10	&	2.43	&	0.95	\\
8	&	2.0	&	2.05	&	1	&	10	&	2.43	&	0.75	\\
8	&	2.1	&	2.02	&	1	&	10	&	2.43	&	0.58	\\
6	&	1.2	&	1.65	&	1	&	10	&	1.80	&	0.63	\\
6	&	1.3	&	1.60	&	1	&	10	&	1.86	&	0.83	\\
6	&	1.4	&	1.56	&	1	&	10	&	1.91	&	0.99	\\
6	&	1.5	&	1.52	&	1	&	10	&	1.95	&	1.1	\\
6	&	1.6	&	1.49	&	1	&	10	&	1.98	&	1.2	\\
6	&	1.7	&	1.46	&	1	&	10	&	2.00	&	1.2	\\
6	&	1.8	&	1.43	&	1	&	10	&	2.02	&	1.2	\\
6	&	1.9	&	1.41	&	1	&	10	&	2.03	&	1.1	\\
6	&	2.0	&	1.39	&	1	&	10	&	2.03	&	0.98	\\
6	&	2.1	&	1.37	&	1	&	10	&	2.03	&	0.86	\\
6	&	2.2	&	1.35	&	1	&	10	&	2.03	&	0.74	\\
6	&	2.3	&	1.33	&	1	&	10	&	2.02	&	0.63	\\
4	&	1.7	&	0.778	&	1	&	10	&	1.57	&	0.55	\\
4	&	1.9	&	0.751	&	1	&	10	&	1.58	&	0.57	\\
4	&	2.1	&	0.729	&	1	&	10	&	1.58	&	0.50	\\
\enddata
\end{deluxetable}

\begin{deluxetable}{cccc}
\tablecaption{Time to reach origin in self-similar collapse, $s_c~=~t_f/t_0~=~dt_f/dt_0$.
Equations \eqref{eq:t0},\eqref{eq:tf} and \eqref{eq:dtfdt0}\label{tbl:sc}}
\tablewidth{0pt} \tablehead{ \colhead{   } & \colhead{$\gamma=4/3$} & \colhead{$1.4$} & \colhead{$5/3$}}
\startdata
	$\der=2$&		$s_c=2.27$	&	2.30 &	2.42	\\
	$3$&		2.09	&	2.17 &	2.21	\\
	$4$&		1.97	&	2.00 &	2.07	\\
\enddata
\end{deluxetable}

\end{document}